# Evaluating tear clearance rate with optical coherence tomography

**Izabela K. Garaszczuk[1], Maryam Mousavi[2], Alejandro Cerviño Expósito[1], Maciej M. Bartuzel[2], Robert Montes-Micó[1] and D. Robert Iskander[2]**

[1]Department of Optics, Optometry and Vision Sciences, University of Valencia, Spain

[2]Department of Biomedical Engineering, Faculty of Fundamental Problem of technology, Wroclaw University of Science and Technology, Poland

Corresponding author at:

I.K. Garaszczuk

Department of Optics, Optometry & Vision Sciences,

University of Valencia,

Calle del Doctor Moliner,

50 - 46100 Burjassot, Spain.

E-mail address: Izabela.garaszczuk@uv.es

# ABSTRACT

**Purpose:** To assess the early-phase of tear clearance rate (TCR) with anterior segment optical coherence tomography (OCT) and to determine the association between TCR and other clinical measures of the tear film in a group of young subjects with different levels of tear film quality.

**Methods:** TCR was classified as the percentage decrease of subject's inferior tear meniscus height 30 seconds after instillation of 5μl 0.9% saline solution. Fifty subjects (32F and 18M) aged (mean ± standard deviation) 25.5 ± 4.3 years volunteered for the study. It consisted of a review of medical history, Ocular Surface Disease Index (OSDI) questionnaire, tear film osmolarity measurements, slit lamp examination and TCR estimation based on dynamic measurements of the lower tear meniscus with OCT. Estimates of TCR were contrasted against subject age and tear film measures commonly used for dry eye diagnosis, which includes OSDI score, fluorescein tear film break-up time (FBUT), tear meniscus height (TMH), blinking frequency, tear film osmolarity and corneal staining.

**Results:** The group mean TCR was 28 ± 13% and 36 ± 19% respectively after 30 and 60 second margin after instillation of saline solution. Statistically significant correlations were found between TCR and FBUT ($r^2=0.319$, $p<0.001$), blinking frequency ($r^2=0.138$, $p<0.01$), tear film osmolarity ($r^2=0.133$, $p<0.01$) and subject's age ($r^2=0.095$, $p<0.05$).

**Conclusions:** Anterior segment optical coherence tomography allows following changes of tear meniscus morphology post saline solution instillation and evaluating the TCR. OCT based TCR might be used as additional measure of the lacrimal functional unit.

**KEYWORDS:** tear turnover, tear clearance, tear meniscus dynamics, optical coherence tomography

**INTRODUCTION**

Tear turnover or tear clearance is described as a global measure of the integrity of the lacrimal functional unit [1–3] and tear exchange on the ocular surface. Tear turnover rate (TTR), a temporal measure of tear turnover is proportional to the sum of the effects of tear secretion by the glands, fluid transudation through the conjunctiva, tear drainage through nasolacrimal duct, evaporation and conjunctival and corneal permeability [2,3] and was shown to be an indirect measure of ocular surface irritation (regardless of reduced or normal aqueous tear production) [4–7], severity of the ocular surface disease [6,7], Meibomian gland dysfunction [6,7] and decreased ocular surface sensitivity [2,5–11]. Also, factors connected with age (conjunctivochalasis, lid laxity, tear flow functional obstruction, blink abnormalities) may all contribute to delayed tear turnover [5,6,12]. Tear Clearance Rate (TCR) is also proven to be reduced in symptomatic dry eye subjects [1,13–15] and in contact lens associated papillary conjunctivitis [16]. Delay in tear clearance can lead to prolonged exposure to topical medications and their preservatives on the ocular surface, thus affected subjects have higher chance to develop ocular surface medication toxicity [5].

The most popular techniques of tear turnover assessment are based on following the elution of tracer molecule added to the tear film with the means of electromagnetic spectrum. This family of methods include fluorophotometry [3,6,12,14,17–27] and lacrimal gamma scintigraphy [28–31]. Fluorophotometry is considered the gold standard in tear turnover and tear flow assessment [27,32]. Standardized procedure, following instillation of 1 µl of 2 % sodium fluorescein into the lower conjunctival sac with a micropipette, lasts up to 30 min [10], when scans are performed every two minutes [1,18] with a commercially available fluorophotometer. In vivo fluorophotometry can be directed to marginal strip [12,18,21] or precorneal tear film [1,3,17–19,21,23–25,33]. The change in rate of fluorescence decay is calculated and tear turnover rate is defined as percentage of fluorescence decay per minute.

Fluorophotometry has some limitations. It requires considerable skill to perform and expensive equipment which lowers its clinical utility. Also extensive period of time is required to obtain results. Due to these limitations, it has been mostly confined to research settings. The aperture of the photometric microscope is larger than the tear film, which is the cause of low spatial resolution of the device [19] that does not allow to measure the fluorescence coming from a thin layer of the tear film without including a portion of the cornea and together with the corneal permeability to sodium fluorescein [3,17,19,25] leads to errors in TTR assessment. Other factors that may contribute to errors in TTR estimation are connected with too high concentrations of fluorescein being instilled [19,21] and counting saturation of the device's electronics [33]. Pearce et al. [24] also pointed out that reduction in blink rate, as well as nonconfluence of the tear film during scans and reflex tearing caused by excessive facial illumination may also contribute to errors in TTR calculation [19,24].

The DEWS I report from 2007 [34] mentioned TTR assessment with fluorophotometry as one of the additional measures of tear film used to diagnose and monitor dry eye disease and addressed the need to develop cheaper and less time consuming methodologies. Recently, a new method for observing tear meniscus morphology by anterior segment optical coherence tomography (AS-OCT) was proposed by Zheng et al. to study the tear clearance as a function of age [35]. Measurements of tear meniscus morphology including tear meniscus height (TMH), depth (TMD), and tear meniscus cross-section area (TMA) have a wide range of applications [36–39]. OCT ensures good repeatability and allows following changes of tear meniscus morphology after fluid instillation [40,41]. In Zheng's study the tear meniscus height, depth and cross-sectional area were measured based on a single scan with in-build OCT software and measurements were repeated 3 times with an interval of at least 15 min between them. In the study presented here, a custom-written algorithm was developed for more precise, automatic estimation of tear meniscus parameters following a blink to simplify the procedure [42]. The aim was to assess the early-phase TCR, utilising an anterior segment OCT and newly-developed software and to compare TCR with tear film measures most commonly used clinically in dry eye diagnosis, to test OCT-based measurements of TCR have potential as a clinically applicable diagnostic method.

**METHODOLOGY**

Fifty healthy subjects (32F and 18M) aged (mean ± standard deviation) 25.5 ± 4.3 years (from 20 to 37) have been recruited for the study. Study adhered to the tenets of the Declaration of Helsinki. Informed consent was obtained from all subjects after the nature and possible consequences of the procedures were explained. Subjects were advised to refrain from wearing contact lenses and instilling ophthalmic solutions at least one week before commencing the study. Exclusion criteria included subjects with signs and symptoms of eye dryness or inflammation, recovering after surgery or with any known tear flow impairment. Slit lamp examination was performed with strong emphasis on sings of lid malformations (entropion or ectropion), conjunctivochalasis (LIPCOF > 1.0) and lacrimal puncta obstruction to ensure that the subjects meet the inclusion criteria.

The study protocol in a chronological order consisted of review of medical history, Ocular Surface Disease Index (OSDI) questionnaire, tear film osmolarity measured with TearLab Osmolarity System (Tear Lab Corp., San Diego, CA, US), slit lamp examination, tear meniscus height, TCR estimation based on inferior tear meniscus height dynamics assessed with anterior segment spectral optical coherence tomography (SOCT Copernicus, Optopol Ltd., Poland), fluorescein tear film break-up time (FBUT) and corneal staining.

The temperature in the laboratory was stable and monitored. The mean temperature was 24.5 ± 1.2 [°C] and mean humidity was 32.2 ± 4.8 [%RH]. Following the evidence that TCR could vary with the daytime [21] measurements were only performed in the morning.

OCT measurements were taken in mesopic conditions to ensure good contrast of the acquired images.

**The assessment of tear meniscus dynamics**

All measurements were performed on the left eye of each of the subjects. Following the procedure presented by Zheng et al. in [35], spectral OCT was used to record the dynamic changes of the inferior tear meniscus. The scanning angle and width was set to 90° and 4 mm, respectively. The B-scan plane, with a maximum of 1800 A-scans, was central (with respect to the iris outline) to the posterior region of the eye-eyelid junction and normal to the eyelid. The maximum possible number of 90 B-scans allowed in that setting was set.

The subject was asked to look straight ahead and refrain from movements. To make sure that the sequence capture the blink period, right after the sequence was initialised, the subject was asked to blink once and then refrain from blinking for the sequence to be finished. Each sequence comprised of 90 frames which corresponds to 3.75 s and can be divided into blink and post-blink intervals. Fig. 1 presents an exemplary B-scan from a sequence.

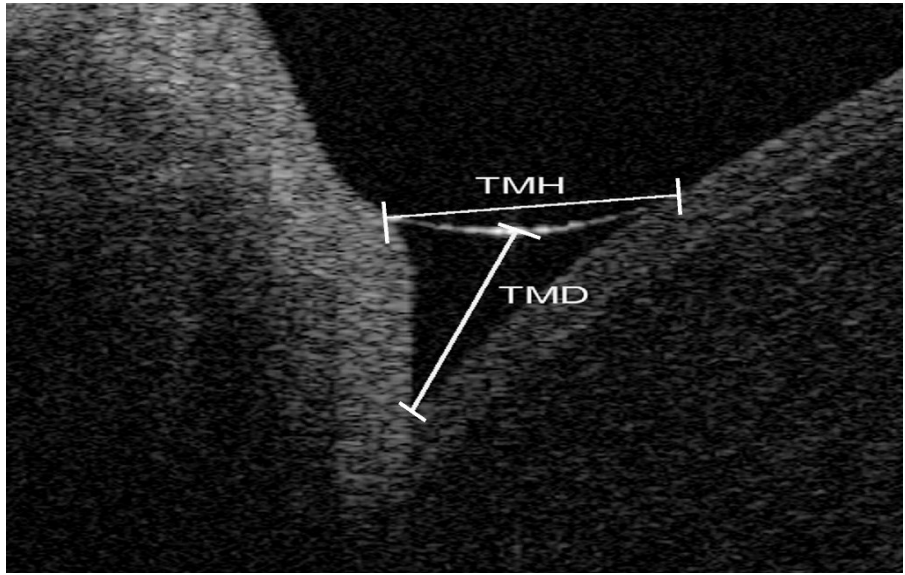

Fig. 1. An exemplary B-scan of the inferior tear meniscus (the image was cropped and resized to show the area of interest), TMH – tear meniscus height, TMD – tear meniscus depth

Baseline tear meniscus morphology of each subject was assessed at the beginning of the procedure, which will be later marked as 'Baseline' (BL) measurement. Subsequently, 5 μl of 0.9% room-temperature saline solution was instilled into the subjects' left eye with a micropipette. Then, shortly after instillation, another sequence was obtained (which is marked as the zero minutes). To follow changes in the tear meniscus morphology over time, the same procedure was repeated after 30 s, 1 min and every minute up to 5 min after the first post-instillation sequence. The TCR was estimated as a percentage decrease in TMH after t = 30 [s]. The reduction in the TMH during the first 30 s post instillation was proven to be the most significant [35].

**Tear Clearance Rate calculations**

Custom-written algorithm was developed [42] to more precisely estimate the tear meniscus height. After a blink the tear meniscus parameters stabilises and it is possible to distinguish frames that correspond to the post-blinking interval. In this interval, the mean value of the inferior TMH was automatically assessed and calculated. Fig. 2 presents all the TMH values from a single 90 B-scan sequence obtained from one of the subjects. With this method one can avoid the influence of post-blink tear meniscus morphology nonconfluence on the acquired data, giving a more precise estimate of the tear meniscus parameters after blink. The software can calculate tear meniscus height, tear meniscus depth and the area of the tear meniscus cross-section. TMH, which was shown to have better clinical utility than tear TMD [35] was used to assess TCR.

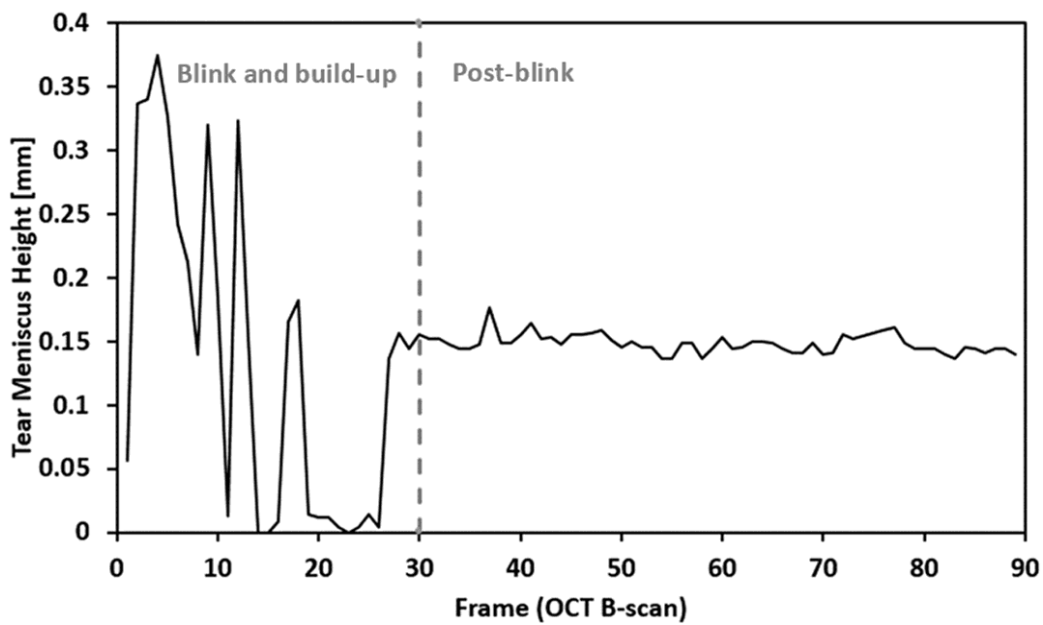

Fig. 2. An exemplary result of TMH dynamics. Dashed grey line indicates an arbitrary division of the TMH temporal changes into a blink and post-blink phase.

**RESULTS**

The group mean TCR was 29 ± 13% and 38 ± 18% at 30 and 60 s margin (range 9% to 57% and 9% to 80%), respectively. The TCR were tested for normality and the null hypothesis was not rejected (Jarque- Bera test, p =0.333). Fig. 3 presents the group mean tear meniscus height dynamics. Statistically significant correlations were found between TCR and FBUT ($r^2 = 0.319$, $p < 0.001$), blinking frequency ($r^2 = 0.135$, $p < 0.01$) and tear film osmolarity ($r^2 = 0.136$, $p < 0.01$). Very low, yet statistically significant correlation was found between TCR and subject's age ($r^2 = 0.095$, $p < 0.05$). No statistically significant correlations were found between TCR and baseline tear meniscus height of the subject, OSDI score and corneal staining. Table 1 shows a summary of the data collected (average values, standard deviations and ranges). Fig. 4 shows all the statistically significant correlations between TCR and other measures assessed in the study.

Table 1. The mean, standard deviation and range of the measured parameters.

| Parameter | Mean ± standard deviation | Range |
|---|---|---|
| Age [years] | 24.5 ± 1.2 | [22, 28] |
| OSDI score | 14.3 ± 12.1 | [0, 32] |
| FTBUT [s] | 14.05 ± 7.96 | [5, 30] |
| Baseline TMH [mm] | 0.26 ± 0.07 | [0.17, 0.41] |
| Blink frequency [$min^{-1}$] | 13 ± 6 | [2, 26] |
| Corneal staining (Efron's Scale) | 0.4 ± 0.4 | [0.0, 1.5] |
| Tear film osmolarity | 301 ± 6 | [289, 318] |
| TCR for 30s and 60s [%] | 29 ± 13 and 38 ± 18 | [9, 57] and [9,80] |

OSDI = Ocular surface disease index questionnaire score, FTBUT = fluorescein tear film break-up time, TMH = tear meniscus height, TCR = tear clearance rate

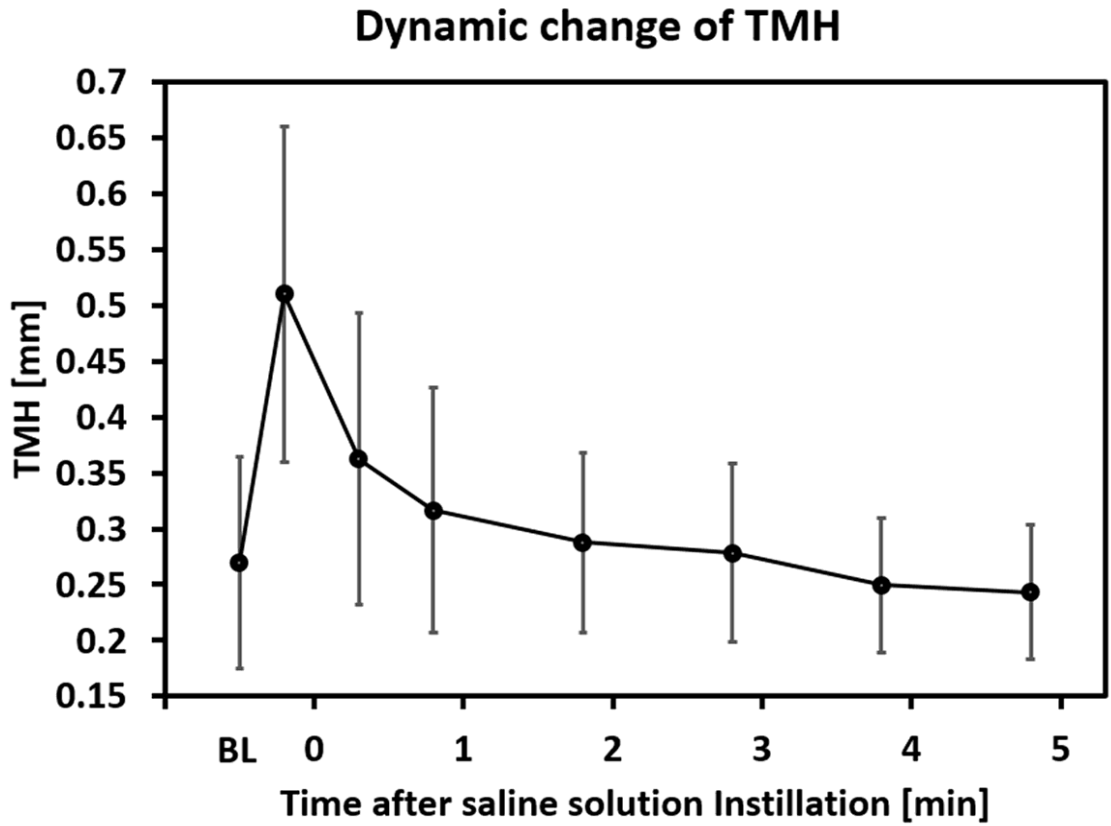

Fig. 3. The group mean TMH changes in time. BL: Baseline - subject's tear meniscus height before the instillation of saline. Error bars indicates ± one standard deviation

.

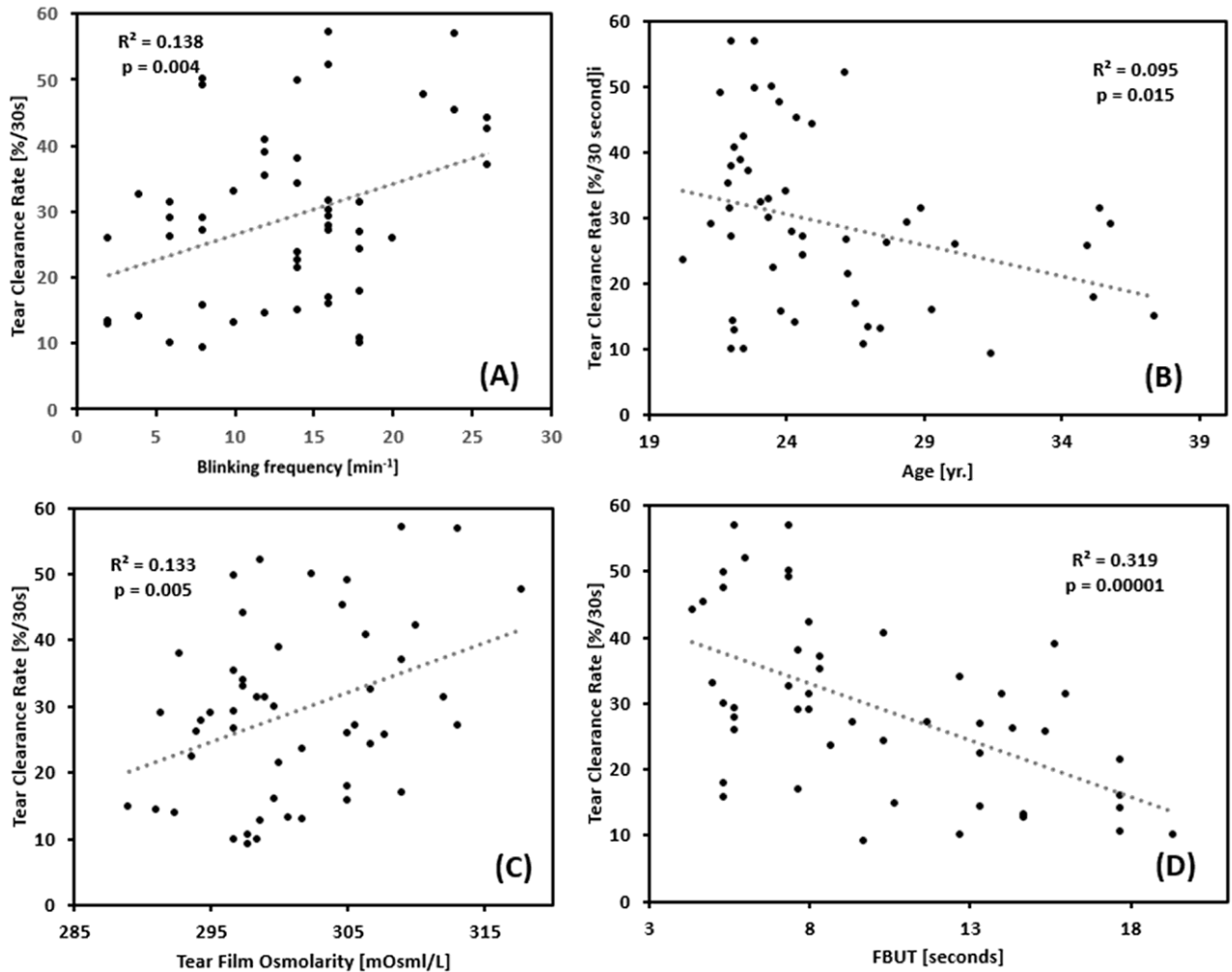

Fig. 4. Statistically significant correlations reported in the study between TCR and (A) blinking frequency, (B) age, (C) tear film osmolarity and (D) fluorescein tear film break-up time.

**Reproducibility**

It has been showed that tear clearance and tear turnover rates differ greatly between subjects [1]. To test the reproducibility of TCR, multiple measurements of dynamic changes in TMH have been performed and TCR has been estimated on a randomly chosen subject for a period of 7 days, with one measurement per day at the same time of the day Fig. 5 shows the dynamic changes in tear meniscus height for that subject. The coefficient of variation of TCR was estimated at 14.85 [%].

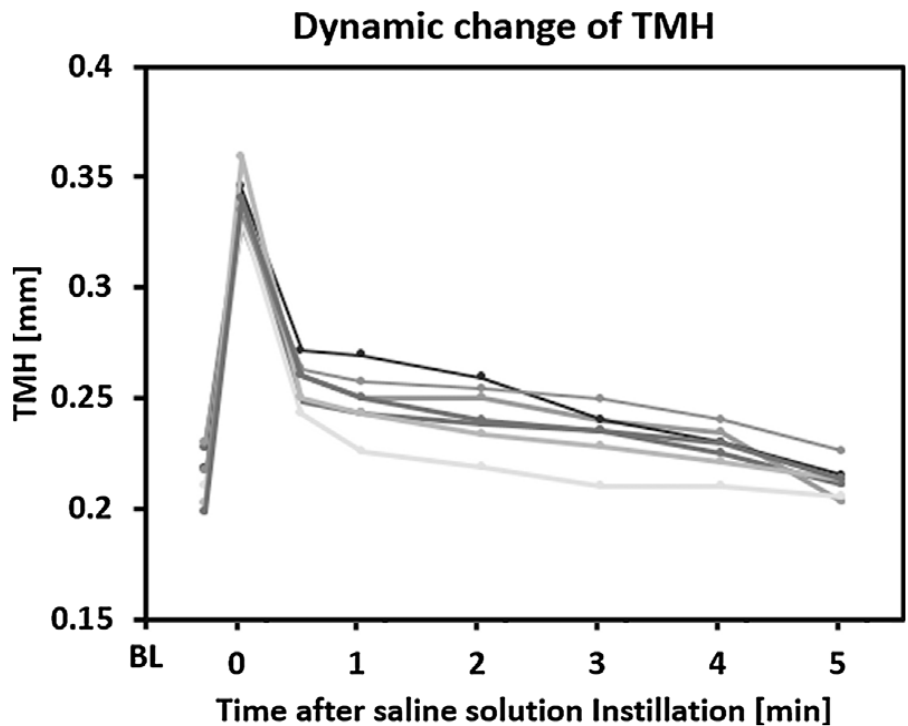

Fig. 5. The dynamic changes of tear meniscus height reported in one subject measured once per day for 7 days.

**DISCUSSION**

The International Dry Eye WorkShop report in 2007 presents TTR assessment with fluorophotometry as one of the additional measures of tear film used to diagnose and monitor dry eye disease and addressed the need to develop cheaper, shorter and more simple methodologies. In this study, we present one of three reported attempts to assess tear clearance rate with optical coherence tomography [35,43]. Since 2014, when the technique was developed, a second study was conducted by the same team utilising Polymethylmethacrylate particles to study the effect of Krehbiel flow. No studies have been made so far by other teams.

This method of TCR estimation is less time consuming, simpler to perform and less invasive than commonly used fluorophotometry.

TMH measurements were taken into consideration in this study to calculate TCR. All images observed were clear enough to be analysed with provided software. In most of the kinetic studies with fluorophotometry, when measurements are made over a longer period of time, the biphasic characteristics of fluorescein clearance are observed [1,3,17–19,22,27], with faster phase occurring just after fluorescein instillation, which seems to be caused by reflex tearing or increased tear volume [5,12,27,44]. The first phase varies with subjects and seems to be correlated with subjective sings of irritation at the time of instillation [27] and seems to be suppressed by anaesthetics [5,12], decreases with age [27] and depends on blinking frequency [12,19,44,45]. The second, slower phase, which occurs usually after five minutes post-instillation presumably represents the tear turnover under basal conditions of secretion and is being used to calculate basal tear turnover rate [18,20].

Findings of Zheng et al. and findings presented in this study suggest that TMH decreased most significantly at early phase following eye drop and shows that TCR estimated with OCT depends on the blink rate and decreases with age. Zheng suggested the effect of Krehbiel flow on the tear clearance rate and showed that TCR assessed with OCT is more likely the manifestation of early-phase tear dynamics than basal tear turnover assessed with fluorophotometry [35,43]. This study further contributes to those developments. It showed that the estimated time for the tear meniscus to come back to its basal level after the instillation of 5 μl of saline solution was about 2–3 min in young healthy subjects, which is in agreement with study of Zheng et al. The reduction of TMH after 30 s post-instillation is most significant.

The baseline value of TMH was $0.26 \pm 0.07$ [mm]. TMH measured immediately after instillation increased significantly to $0.51 \pm 0.15$ [mm] ($p < 0.0001$). Following 30 s of natural blinking, TMH decreased significantly compared to 0 s ($p < 0.001$). Changes between 2 and 3 min (and later) were not statistically insignificant (2 and 3 min, p = 0.143; 3 and 4 min, p =0.06 and 4 and 5 min, p = 0.111). The mean tear clearance rate was estimated as 29 ± 13 [%] compared to 35 ± 11 [%] in the younger group reported in Zheng et al. and also showed large variation between subjects, with relatively good reproducibility.

In Zheng's study TMH clearance rates were significantly and negatively correlated with the degree of conjunctivochalasis, the distance between the inferior lacrimal punctum and Marx line, and the degree of protrusion of the lacrimal punctum. The study presented here suggests that TCR can be also correlated with more clinically applicable tear film measures, such as the tear film break-up time ($r^2 = 0.319$, $p < 0.001$) and tear film osmolarity ($r^2 = 0.133$, $p < 0.01$). In both studies of Zheng et al. the age-related differences in TCR were evident, noting a wide age gap between tested groups of subjects (35.2 ± 11% versus 12.4 ± 7.3% for younger [29.6 ± 7.2 y.o.] and older group [71.4 ± 10.8 y.o.], respectively). This study made attempt to test the improved OCT methodology on a group of subjects with a more narrow age range. The correlation between TCR and age is not strong ($r^2 = 0.095$), however it was found to be statistically significant ($p < 0.05$). Increasing the age range would most definitely improve that correlation.

The proposed use of the implemented custom-written software allows more control over the tear meniscus height following each blink, to observe the moment when tear meniscus height stabilises over time and delete outliers from the final TMH estimation. It takes into account small eye movements and continuous changes of TMH after each blink, that cannot be observed based on a single image. The inferior TMH is not constant but increases with time after a blink [46]. Right after the blink, the rate of change is very rapid, resulting in a relatively stable TMH after a short time and small standard deviation, which could be also observed with the software. An additional functionality that should be considered in the future studies of tear clearance with OCT is to control the amount of blinks between each measurements or to assess the tear meniscus height in a continuous manner, as it was observed by Wu et al. [47]. In various studies it was shown, that the blink rate influences tear turnover [19], and hence could be considered as a limitation. The results show statistically significant correlation between TCR measured with OCT and blink frequency assessed in normal conditions. It would be also of interest to differentiate the full blinks from partial ones. Controlled blinking frequency may disturb the natural process of tear exchange. However, taking into consideration its expected impact on the results it would be of interest to measure it simultaneously with TCR or to control it. Future studies should investigate applicability of these techniques on dry eye subjects as well as the qualitative differences in early-phase tear dynamics, as it is shown that tear turnover assessed with fluorophotometry is reduced in subjects with symptomatic dry eye [13]. The 3.75 s' period contains the sequence that corresponds to the blink and post blink period, so the time, when the subject eye is open is usually less than 3s. In all subjects this amount of time was not associated with any observable effort. Despite the statistical significance the $r^2$ values for TCR are low, but we know that it is not an uncommon feature of commonly used test of tear film dynamics [48].

Summarising, OCT can be used as a rapid, qualitative and quantitative method of determining tear clearance rate. With the new algorithm developed, tear meniscus parameters can be calculated more precisely, considering the nonconfluence of tear meniscus morphology following each blink and small eye movements. Traditional clearance tests are either invasive or laborious, and the use of fluorescein has some limitation, because the clearance of the tear fluids is not directly measured. Tear clearance measured with OCT is non-invasive and relatively more rapid and simpler to perform that the traditionally used tear clearance tests. TCR measurements also suggest that in healthy subjects the tear meniscus height comes back to its basal state 2 to 3 min after fluid instillation. This or similar technique may be useful in testing artificial tears performance [49,50]. TCR is proven to decrease in elderly subjects and symptomatic subjects. TCR also correlates with age. Apparent, statistically significant, yet small correlation reported in this study may be due to a narrow age group measured, which was comprised mostly of young, healthy individuals. Newly developed algorithm allows precise, automatic estimation of tear meniscus parameters. Presented correlations of TCR with tear film osmolarity and tear film break up time could lead to a better understanding of the complexity of tear film dynamics and the role of tear clearance in the pathogenesis of dry eye syndrome and ocular irritation.

## CONFLICT OF INTEREST

None.

## ACKNOWLEDGEMENTS

This project has received funding from the European Union's Horizon 2020 research and innovation programme under the Marie Sklodowska-Curie grant agreement No 642760. Part of this work has been presented at the British Contact Lens Society 40th Clinical Conference & Exhibition, Liverpool, June 2017.